# An action with positive kinetic energy term for general relativity

T. Mei

(Department of Journal, Central China Normal University, Wuhan, Hubei PRO, People's Republic of China
E-Mail:   meitao@mail.ccnu.edu.cn    meitaowh@public.wh.hb.cn )

**Abstract:**   At first, we state some results in arXiv: 0707.2639, and then, using a positive kinetic energy coordinate condition given by arXiv: 0707.2639, we present an action with positive kinetic energy term for general relativity. Based on this action, the corresponding theory of canonical quantization is discussed.

## 1   Some results in arXiv: 0707.2639

### 1.1   Basic form of the theory

Considering the tetrad field $e_\mu^{\hat\alpha}$ by which the metric tensor $g_{\mu\nu}$ is

$$g_{\mu\nu} = \eta_{\hat\alpha\hat\beta} e_\mu^{\hat\alpha} e_\nu^{\hat\beta} = e_\mu^{\hat\alpha} e_{\hat\alpha\nu} . \tag{1}$$

Define

$$F_{\mu\nu\lambda} = r_{\hat\alpha\hat\beta\hat\gamma} e_\mu^{\hat\alpha} e_\nu^{\hat\beta} e_\lambda^{\hat\gamma} = e_\mu^{\hat\alpha} e_{\hat\alpha\nu;\lambda}$$
$$= \frac{1}{2} e_\mu^{\hat\alpha} \left( e_{\hat\alpha\nu,\lambda} - e_{\hat\alpha\lambda,\nu} \right) - \frac{1}{2} e_\nu^{\hat\alpha} \left( e_{\hat\alpha\mu,\lambda} - e_{\hat\alpha\lambda,\mu} \right) - \frac{1}{2} e_\lambda^{\hat\alpha} \left( e_{\hat\alpha\mu,\nu} - e_{\hat\alpha\nu,\mu} \right),$$

where $r_{\hat\alpha\hat\beta\hat\gamma}$ is the Ricci's coefficients of rotation:

$$r_{\hat\alpha\hat\beta\hat\gamma} = e_{\hat\alpha\mu;\nu} e_{\hat\beta}^\mu e_{\hat\gamma}^\nu = \frac{1}{2}\left( e_{\hat\alpha\mu,\nu} - e_{\hat\alpha\nu,\mu} \right) e_{\hat\beta}^\mu e_{\hat\gamma}^\nu - \frac{1}{2}\left( e_{\hat\beta\mu,\nu} - e_{\hat\beta\nu,\mu} \right) e_{\hat\alpha}^\mu e_{\hat\gamma}^\nu - \frac{1}{2}\left( e_{\hat\gamma\mu,\nu} - e_{\hat\gamma\nu,\mu} \right) e_{\hat\alpha}^\mu e_{\hat\beta}^\nu .$$

The Einstein-Hilbert action

$$S_{\text{EH}} = \frac{c^3}{16\pi G} \int \sqrt{-g}\, d^4x\, R = \frac{c^3}{16\pi G} \int \left|{}^4 e\right| d^4 x\, L_{\text{G}} + \frac{c^3}{16\pi G} \int d^4 x \frac{\partial}{\partial x^\rho}(-2\sqrt{-g}\, F^{\rho\sigma}{}_\sigma),$$

$$L_{\text{G}} = F^{\rho\sigma}{}_\sigma F_{\rho\lambda}{}^\lambda - F^{\rho\sigma\lambda} F_{\rho\lambda\sigma} = r^{\hat\alpha\hat\beta}{}_{\hat\alpha} r^{\hat\gamma}{}_{\hat\beta\hat\gamma} - r^{\hat\alpha\hat\beta\hat\gamma} r_{\hat\alpha\hat\gamma\hat\beta}$$

$$= -\frac{1}{4}\left( \eta_{\hat\alpha\hat\beta} g^{\mu\rho} g^{\nu\sigma} + 2 g^{\mu\rho} e_{\hat\alpha}^\sigma e_{\hat\beta}^\nu - 4 g^{\mu\rho} e_{\hat\alpha}^\nu e_{\hat\beta}^\sigma \right)\left( e_{\mu,\nu}^{\hat\alpha} - e_{\nu,\mu}^{\hat\alpha} \right)\left( e_{\rho,\sigma}^{\hat\beta} - e_{\sigma,\rho}^{\hat\beta} \right). \tag{2}$$

The corresponding Euler-Lagrange equations

$$\frac{\partial\left[\left|{}^4 e\right|\left(\frac{c^3}{16\pi G} L_{\text{G}} + L_{\text{M}}\right)\right]}{\partial e_{\hat\alpha\mu}} - \partial_\nu \frac{\partial\left[\left|{}^4 e\right|\left(\frac{c^3}{16\pi G} L_{\text{G}} + L_{\text{M}}\right)\right]}{\partial e_{\hat\alpha\mu,\nu}} \equiv -\frac{c^3 \left|{}^4 e\right|}{8\pi G} \Theta^{\mu\hat\alpha} = 0,$$

$$\Theta^{\mu\hat\alpha} = \frac{1}{\left|{}^4 e\right|} \frac{\partial}{\partial x^\nu}\left(\left|{}^4 e\right| S^{\mu\nu\hat\alpha}\right) + e^{\hat\alpha\rho} S^{\mu\sigma\hat\beta}\left( e_{\hat\beta\rho,\sigma} - e_{\hat\beta\sigma,\rho} \right) - \frac{1}{2} e^{\hat\alpha\mu} L_{\text{G}} - \frac{8\pi G}{c^4} T^{\mu\hat\alpha},$$

$$S^{\mu\nu\hat{\alpha}} = S^{\mu\nu\lambda}e_\lambda^{\hat{\alpha}}$$
$$= \frac{1}{2}\left(e_{\hat{\beta}}^\mu g^{\nu\rho} - e_{\hat{\beta}}^\nu g^{\mu\rho}\right)e^{\hat{\alpha}\sigma}\left(e_{\rho,\sigma}^{\hat{\beta}} - e_{\sigma,\rho}^{\hat{\beta}}\right)$$
$$-\left(e^{\hat{\alpha}\mu}g^{\nu\rho} - e^{\hat{\alpha}\nu}g^{\mu\rho}\right)e_{\hat{\beta}}^\sigma\left(e_{\rho,\sigma}^{\hat{\beta}} - e_{\sigma,\rho}^{\hat{\beta}}\right) - \frac{1}{2}g^{\mu\rho}g^{\nu\sigma}\left(e_{\rho,\sigma}^{\hat{\alpha}} - e_{\sigma,\rho}^{\hat{\alpha}}\right).$$

There are six identities and ten independent equations being equivalent to the Einstein equations in the sixteen Euler-Lagrange equations.

### 1.2 The Schwinger time gauge condition

$$e_{\hat{a}}^0 = 0, \quad a = 1, 2, 3, \tag{3}$$

Under the condition (3), (2) is simplified to the following form

$$L_G = L_{G0} + 2e_{\hat{0}}^0 e_{0,k}^{\hat{0}} U^k + L_{GV},$$
$$L_{G0} = \frac{1}{2}M_{\hat{a}\hat{b}}^{ij} e_{\hat{0}}^\mu \left(e_{i,\mu}^{\hat{a}} - e_{\mu,i}^{\hat{a}}\right)e_{\hat{0}}^\nu \left(e_{j,\nu}^{\hat{b}} - e_{\nu,j}^{\hat{b}}\right) = \left(e_{\hat{0}}^0\right)^2 \overline{L}_{G0},$$
$$U^k = e_{\hat{a}}^i e_{\hat{b}}^j e^{\hat{b}k}\left(e_{i,j}^{\hat{a}} - e_{j,i}^{\hat{a}}\right),$$
$$L_{GV} = e^{\hat{c}l}e_{\hat{c}}^m\left(-\frac{1}{4}\eta_{\hat{a}\hat{b}}e^{\hat{d}i}e_{\hat{d}}^j - \frac{1}{2}e_{\hat{a}}^j e_{\hat{b}}^i + e_{\hat{a}}^i e_{\hat{b}}^j\right)\left(e_{i,l}^{\hat{a}} - e_{l,i}^{\hat{a}}\right)\left(e_{j,m}^{\hat{b}} - e_{m,j}^{\hat{b}}\right), \tag{4}$$
$$M_{\hat{a}\hat{b}}^{ij} = \eta_{\hat{a}\hat{b}}e^{\hat{c}i}e_{\hat{c}}^j + e_{\hat{a}}^j e_{\hat{b}}^i - 2e_{\hat{a}}^i e_{\hat{b}}^j,$$
$$\overline{L}_{G0} = \frac{1}{2}M_{\hat{a}\hat{b}}^{ij}\left(e_{i,0}^{\hat{a}} - e_{0,i}^{\hat{a}}\right)\left(e_{j,0}^{\hat{b}} - e_{0,j}^{\hat{b}}\right) - M_{\hat{a}\hat{b}}^{ij}\left(e_{i,0}^{\hat{a}} - e_{0,i}^{\hat{a}}\right)e_{\hat{d}}^0 e^{\hat{d}m}\left(e_{j,m}^{\hat{b}} - e_{m,j}^{\hat{b}}\right)$$
$$+ \frac{1}{2}M_{\hat{a}\hat{b}}^{ij}e_{\hat{0}}^{\hat{c}}e_{\hat{c}}^l\left(e_{i,l}^{\hat{a}} - e_{l,i}^{\hat{a}}\right)e_{\hat{0}}^{\hat{d}}e_{\hat{d}}^m\left(e_{j,m}^{\hat{b}} - e_{m,j}^{\hat{b}}\right).$$

In (4), time derivative term only appears in $L_{G0}$ and there is not the term $e_{0,0}^{\hat{\alpha}}$ in $L_{G0}$.

The sixteen Euler-Lagrange equations $\Theta^{\mu\hat{\alpha}} = 0$ under the condition (3) read

$$-2e_{\hat{0}}^{\hat{0}}\Theta^{0\hat{0}} = -2\Theta^{\hat{0}\hat{0}} = L_{G0} + \frac{2}{|{}^3e|}\frac{\partial}{\partial x^k}\left(|{}^3e|U^k\right) - L_{GV} + \frac{16\pi G}{c^4}T^{\hat{0}\hat{0}} = 0, \tag{5}$$

$$e_{\hat{0}}^{\hat{0}}\Theta^{0\hat{a}} = \Theta^{\hat{0}\hat{a}} = \frac{1}{|{}^3e|}\frac{\partial}{\partial x^k}\left(|{}^3e|S^{\hat{0}k\hat{a}}\right) + e^{\hat{a}i}S^{\hat{0}j\hat{b}}\left(e_{\hat{b}i,j} - e_{\hat{b}j,i}\right) - \frac{8\pi G}{c^4}T^{\hat{0}\hat{a}} = 0, \tag{6}$$

where

$$S^{\hat{0}i\hat{a}} = e_{\hat{0}}^{\hat{0}}S^{0i\hat{a}} = e^{\hat{a}i}e_{\hat{0}}^\lambda e_{\hat{b}}^j\left(e_{j,\lambda}^{\hat{b}} - e_{\lambda,j}^{\hat{b}}\right) - \frac{1}{2}e_{\hat{b}}^i e_{\hat{0}}^\lambda\left[e^{\hat{b}j}\left(e_{j,\lambda}^{\hat{a}} - e_{\lambda,j}^{\hat{a}}\right) + e^{\hat{a}j}\left(e_{j,\lambda}^{\hat{b}} - e_{\lambda,j}^{\hat{b}}\right)\right]. \tag{7}$$

Because there is not $e_{\hat{0}i}^{\hat{0}}$ in $e_{\hat{0}}^{\hat{0}}|{}^3e|L_G$, there is not the corresponding equation $\Theta^{i\hat{0}} = 0$ in the Euler-Lagrange equations. On the other hand, we can prove

$$\Theta^{i\hat{0}} = \Theta^{i\mu}e_\mu^{\hat{0}} = \Theta^{i0}e_0^{\hat{0}} = \Theta^{0i}e_0^{\hat{0}} = e_{\hat{a}}^i\left(\Theta^{0\hat{a}}e_{\hat{0}}^0 - \Theta^{0\hat{0}}e_0^{\hat{a}}\right),$$

It shows that $\Theta^{i\hat{0}} = 0$ does not provide new independent equation.

The rest nine equations are

$$e_{\hat{0}}^{\hat{0}}\Theta^{i\hat{a}} = -\frac{1}{|^3e|}\frac{\partial}{\partial x^0}\left(|^3e|S^{\hat{0}i\hat{a}}\right) + \frac{1}{|^3e|}\frac{\partial}{\partial x^j}\left[|^3e|\left(e_{\hat{0}}^{\hat{b}}e_{\hat{b}}^{j}S^{\hat{0}i\hat{a}} - e_{\hat{0}}^{\hat{b}}e_{\hat{b}}^{i}S^{\hat{0}j\hat{a}} + e_{\hat{0}}^{\hat{0}}\tilde{s}^{ij\hat{a}}\right)\right]$$
$$+ e_{\hat{0},j}^{\hat{0}}e^{\hat{a}j}U^i - e^{\hat{a}j}S^{\hat{0}i\hat{b}}\left(e_{\hat{b}j,0} - e_{\hat{b}0,j}\right)$$
$$+ e^{\hat{a}j}\left(e_{\hat{0}}^{\hat{c}}e_{\hat{c}}^{k}S^{\hat{0}i\hat{b}} - e_{\hat{0}}^{\hat{c}}e_{\hat{c}}^{i}S^{\hat{0}k\hat{b}} + e_{\hat{0}}^{\hat{0}}\tilde{s}^{ik\hat{b}}\right)\left(e_{\hat{b}j,k} - e_{\hat{b}k,j}\right) - \frac{8\pi G}{c^4}e_{\hat{0}}^{\hat{0}}T^{i\hat{a}} = 0,$$
(8)

where $U^i$ and $S^{\hat{0}i\hat{a}}$ are given by (4) and (7), respectively;

$$\tilde{s}^{ij\hat{a}} = e_{\hat{0}}^{0}e_{0,k}^{\hat{0}}e_{\hat{b}}^{k}\left(e^{\hat{a}i}e^{\hat{b}j} - e^{\hat{b}i}e^{\hat{a}j}\right) - \frac{1}{2}e_{\hat{b}}^{i}e^{\hat{b}l}e_{\hat{c}}^{j}e^{\hat{c}m}\left(e_{l,m}^{\hat{a}} - e_{m,l}^{\hat{a}}\right)$$
$$+ \frac{1}{2}e^{\hat{a}l}e^{\hat{b}m}\left(e_{\hat{b}}^{i}e_{\hat{c}}^{j} - e_{\hat{b}}^{j}e_{\hat{c}}^{i}\right)\left(e_{l,m}^{\hat{c}} - e_{m,l}^{\hat{c}}\right) - \left(e^{\hat{a}i}e^{\hat{b}j} - e^{\hat{b}i}e^{\hat{a}j}\right)e_{\hat{b}}^{l}e_{\hat{c}}^{m}\left(e_{l,m}^{\hat{c}} - e_{m,l}^{\hat{c}}\right).$$

### 1.3 The non-positive definiteness of the quadratic term of time derivative in $L_G$

The quadratic term of time derivative in $L_{G0}$ given by (4) is non-positive definitive. This conclusion is obvious from the following expression:

$$L_{G0} = -\frac{2}{3}\left[e_{\hat{0}}^{\lambda}e_{\hat{a}}^{i}\left(e_{i,\lambda}^{\hat{a}} - e_{\lambda,i}^{\hat{a}}\right)\right]^2 + \frac{1}{6}\left\{e_{\hat{0}}^{\lambda}\left[2e^{\hat{1}i}\left(e_{i,\lambda}^{\hat{1}} - e_{\lambda,i}^{\hat{1}}\right) - e^{\hat{2}i}\left(e_{i,\lambda}^{\hat{2}} - e_{\lambda,i}^{\hat{2}}\right) - e^{\hat{3}i}\left(e_{i,\lambda}^{\hat{3}} - e_{\lambda,i}^{\hat{3}}\right)\right]\right\}^2$$
$$+ \frac{1}{2}\left\{e_{\hat{0}}^{\lambda}\left[e^{\hat{2}i}\left(e_{i,\lambda}^{\hat{2}} - e_{\lambda,i}^{\hat{2}}\right) - e^{\hat{3}i}\left(e_{i,\lambda}^{\hat{3}} - e_{\lambda,i}^{\hat{3}}\right)\right]\right\}^2 + \frac{1}{2}\left\{e_{\hat{0}}^{\lambda}\left[e^{\hat{1}i}\left(e_{i,\lambda}^{\hat{2}} - e_{\lambda,i}^{\hat{2}}\right) + e^{\hat{2}i}\left(e_{i,\lambda}^{\hat{1}} - e_{\lambda,i}^{\hat{1}}\right)\right]\right\}^2 \quad (9)$$
$$+ \frac{1}{2}\left\{e_{\hat{0}}^{\lambda}\left[e^{\hat{1}i}\left(e_{i,\lambda}^{\hat{3}} - e_{\lambda,i}^{\hat{3}}\right) + e^{\hat{3}i}\left(e_{i,\lambda}^{\hat{1}} - e_{\lambda,i}^{\hat{1}}\right)\right]\right\}^2 + \frac{1}{2}\left\{e_{\hat{0}}^{\lambda}\left[e^{\hat{2}i}\left(e_{i,\lambda}^{\hat{3}} - e_{\lambda,i}^{\hat{3}}\right) + e^{\hat{3}i}\left(e_{i,\lambda}^{\hat{2}} - e_{\lambda,i}^{\hat{2}}\right)\right]\right\}^2.$$

$L_{G0}$ can be written to the following form:

$$L_{G0} = e_{\hat{0}}^{\mu}e_{\hat{0}}^{\nu}\left(\bar{g}^{il}\bar{g}^{jm} - \bar{g}^{ij}\bar{g}^{lm}\right)\Gamma_{\mu ij}\Gamma_{vlm},$$
(10)

where $\Gamma_{\lambda ij} = \frac{1}{2}\left(g_{\lambda i,j} + g_{\lambda j,i} - g_{ij,\lambda}\right)$, $\bar{g}^{ij} = e_{\hat{a}}^{i}e^{\hat{a}j} = g^{ij} + e_{\hat{0}}^{i}e_{\hat{0}}^{j}$, and $\bar{g}^{ik}g_{kj} = \delta_{j}^{i}$.

### 1.4 A coordinate condition insuring positive definiteness of the kinetic energy term in $L_{G0}$

The following formula is proved:

$$e_{\hat{0}}^{\lambda}e_{\hat{a}}^{i}\left(e_{i,\lambda}^{\hat{a}} - e_{\lambda,i}^{\hat{a}}\right) = \frac{1}{\sqrt{-g}}\left(\sqrt{|g_{lm}|}\frac{g^{0\lambda}}{g^{00}}\right)_{,\lambda} = \sqrt{-g^{00}}\left[\frac{1}{2}\frac{|g_{lm}|_{,0}}{|g_{lm}|} + \frac{1}{2}\frac{g^{0i}}{g^{00}}\frac{|g_{lm}|_{,i}}{|g_{lm}|} + \left(\frac{g^{0i}}{g^{00}}\right)_{,i}\right]. \quad (11)$$

From (11) we see that if we choose

$$\left(\sqrt{|g_{lm}|}\frac{g^{0\lambda}}{g^{00}}\right)_{,\lambda} = 0,$$
(12)

then

$$e_{\hat{0}}^{\lambda}e_{\hat{a}}^{i}\left(e_{i,\lambda}^{\hat{a}} - e_{\lambda,i}^{\hat{a}}\right) = 0,$$
(13)

and the quadratic term of time derivative in $L_{G0}$ given by (9) is thus positive definitive.

### 1.5 A group of variable substitution

Using (3) and (1) we have

$$e_i^{\hat{0}} = 0,\ i=1,2,3;\ e_{\hat{0}}^0 = \left(e_0^{\hat{0}}\right)^{-1};\ e_{\hat{0}}^i = -e_{\hat{0}}^0 e_{\hat{a}}^i e_0^{\hat{a}};\ \sqrt{-g} = \left|{}^4 e\right| = e_0^{\hat{0}}\left|{}^3 e\right|;\ e_{\hat{a}}^i e_i^{\hat{b}} = \delta_{\hat{a}}^{\hat{b}};\ e_{\hat{a}}^j e_i^{\hat{a}} = \delta_i^j,\quad (14)$$

$$g^{00} = -\left(e_{\hat{0}}^0\right)^2,\ g^{0i} = -e_{\hat{0}}^0 e_{\hat{0}}^i,\ g^{ij} = -e_{\hat{0}}^i e_{\hat{0}}^j + e_{\hat{a}}^i e^{\hat{a}j};$$

$$g_{00} = -\left(e_0^{\hat{0}}\right)^2 + e_0^{\hat{a}} e_{\hat{a}0},\ g_{0i} = e_0^{\hat{a}} e_{\hat{a}i},\ g_{ij} = e_i^{\hat{a}} e_{\hat{a}j}.\quad (15)$$

$$g_{33} = \sum_{a=1}^{3}\left(e_3^{\hat{a}}\right)^2 > 0,\ \begin{vmatrix} g_{22} & g_{23} \\ g_{32} & g_{33} \end{vmatrix} = \left(e_2^{\hat{1}} e_3^{\hat{2}} - e_3^{\hat{1}} e_2^{\hat{2}}\right)^2 + \left(e_2^{\hat{1}} e_3^{\hat{3}} - e_3^{\hat{1}} e_2^{\hat{3}}\right)^2 + \left(e_2^{\hat{2}} e_3^{\hat{3}} - e_3^{\hat{2}} e_2^{\hat{3}}\right)^2 > 0,$$

$$\left|g_{ij}\right| = \left|{}^3 e\right|^2 > 0.\quad (16)$$

Based on (16), we define

$$\sqrt{\left|g_{ij}\right|} = h_0,\ \frac{\sqrt[4]{\left[g_{22} g_{33} - (g_{23})^2\right]^3}}{\sqrt{\left|g_{ij}\right|}} = h_1,\ \frac{\sqrt{g_{33}}}{\sqrt[4]{g_{22} g_{33} - (g_{23})^2}} = h_2,$$

$$h_3 = \frac{g_{23}}{g_{33}},\ h_4 = \frac{g_{23} g_{31} - g_{12} g_{33}}{g_{22} g_{33} - (g_{23})^2},\ h_5 = \frac{g_{12} g_{23} - g_{22} g_{31}}{g_{22} g_{33} - (g_{23})^2}.\quad (17)$$

Taking advantage of (17), $\sqrt{\left|g_{ij}\right|}$ as an independent variable is separated from the six dynamical variables $g_{ij}$. Contrarily, we can obtain easily the expression

$$g_{ij} = g_{ij}(h_u),\ u=0,1,2,3,4,5 \quad (18)$$

from (17). And, if there does not exist gravitation field, and $g_{11} = g_{22} = g_{33} = 1$, $g_{12} = g_{23} = g_{31} = 0$ (Minkowski metric), then $h_0 = h_1 = h_2 = 1,\ h_3 = h_4 = h_5 = 0$.

**1.6 The Cholesky decomposition**

For the purpose that $e_i^{\hat{a}}$ can be expressed by $g_{ij}$, and, further, by $h_u$ ($u$=0, 1, 2, 3, 4, 5) via (18):

$$e_i^{\hat{a}} = e_i^{\hat{a}}(g_{lm}) = e_i^{\hat{a}}(h_u),$$

we add following gauge conditions

$$e_{\hat{\alpha}\mu} = 0,\ \alpha < \mu,\quad (19)$$

combining the last formula in (15), $e_{\hat{a}i}$ is thus a triangular matrix:

$$\begin{bmatrix} e_{\hat{1}1} & e_{\hat{1}2} & e_{\hat{1}3} \\ e_{\hat{2}1} & e_{\hat{2}2} & e_{\hat{2}3} \\ e_{\hat{3}1} & e_{\hat{3}2} & e_{\hat{3}3} \end{bmatrix} = \begin{bmatrix} \dfrac{\sqrt{\left|g_{ij}\right|}}{\sqrt{g_{22} g_{33} - (g_{23})^2}} & 0 & 0 \\ -\dfrac{g_{23} g_{31} - g_{12} g_{33}}{\sqrt{g_{33}}\sqrt{g_{22} g_{33} - (g_{23})^2}} & \dfrac{\sqrt{g_{22} g_{33} - (g_{23})^2}}{\sqrt{g_{33}}} & 0 \\ \dfrac{g_{31}}{\sqrt{g_{33}}} & \dfrac{g_{23}}{\sqrt{g_{33}}} & \sqrt{g_{33}} \end{bmatrix}\quad (20)$$

$$= \begin{bmatrix} \sqrt[3]{h_0 h_1} \dfrac{1}{h_1} & 0 & 0 \\ -\sqrt[3]{h_0 h_1} \dfrac{h_4}{h_2} & \sqrt[3]{h_0 h_1} \dfrac{1}{h_2} & 0 \\ -\sqrt[3]{h_0 h_1} h_2 (h_3 h_4 + h_5) & \sqrt[3]{h_0 h_1} h_2 h_3 & \sqrt[3]{h_0 h_1} h_2 \end{bmatrix}. \quad (21)$$

The form of $e_{\hat{a}i}$ given by (20) is so called Cholesky decomposition in algebra.

Based on (20), from $g_{0i} = e_0^{\hat{a}} e_{\hat{a}i}$ in (15) we have

$$e_{\hat{1}0} = \dfrac{g_{01}[g_{22}g_{33} - (g_{23})^2] + g_{02}(g_{23}g_{31} - g_{12}g_{33}) + g_{03}(g_{12}g_{23} - g_{22}g_{31})}{\sqrt{|g_{ij}|}\sqrt{g_{22}g_{33} - (g_{23})^2}},$$

$$e_{\hat{2}0} = \dfrac{g_{02}g_{33} - g_{03}g_{23}}{\sqrt{g_{33}}\sqrt{g_{22}g_{33} - (g_{23})^2}}, \quad e_{\hat{3}0} = \dfrac{g_{03}}{\sqrt{g_{33}}}. \quad (22)$$

$$e_0^{\hat{0}} = \left(e_{\hat{0}}^0\right)^{-1} = \sqrt{-g^{00}}, \quad e_0^i = \dfrac{-g^{0i}}{\sqrt{-g^{00}}}. \quad (23)$$

## 2  An action with positive kinetic energy term for general relativity

In §1, we state some results in arXiv:0707.2639, one among the results is that the quadratic term of time derivative in $L_{G0}$ given by (4) is non-positive definitive; but, generally speaking, the quadratic term of time derivative in an action corresponds to kinetic energy of the system, if this term was non-positive, then it was weird. On the other hand, the non-positive definiteness of the quadratic term of time derivative in an action leads to the principle of variation failure[2].

On the other hand, yet based on the results given by §1, we can present an action with positive kinetic energy term for general relativity.

Substitute (13) into (5), (6) and (8), we obtain *the Einstein equations with the character (13) under the condition (3)*, whose concrete forms no longer be written down here.

In (9), the negative kinetic energy term in $L_{G0}$ is

$$L_{GNK} = -\dfrac{2}{3}\left[e_0^{\lambda} e_{\hat{a}}^i \left(e_{i,\lambda}^{\hat{a}} - e_{\lambda,i}^{\hat{a}}\right)\right]^2 = \dfrac{2}{3g}\left[\left(\sqrt{|g_{lm}|}\dfrac{g^{0\lambda}}{g^{00}}\right)_{,\lambda}\right]^2$$

$$= -\dfrac{2}{3}\left\{\sqrt{-g^{00}}\left[\dfrac{1}{2}\dfrac{|g_{lm}|_{,0}}{|g_{lm}|} + \dfrac{1}{2}\dfrac{g^{0i}}{g^{00}}\dfrac{|g_{lm}|_{,i}}{|g_{lm}|} + \left(\dfrac{g^{0i}}{g^{00}}\right)_{,i}\right]\right\}^2 \quad (24)$$

$$= \dfrac{2}{3}g^{00}\left[\dfrac{1}{2}g^{lm}g_{lm,0} + \dfrac{1}{2}\dfrac{g^{0i}}{g^{00}}g^{lm}g_{lm,i} + \left(\dfrac{g^{0i}}{g^{00}}\right)_{,i}\right]^2;$$

It is important that we can prove that the action

$$S = \dfrac{c^3}{16\pi G}\int |{}^4e| \mathrm{d}^4 x\, L_{GPK}, \quad (25)$$

$$L_{GPK} = L_G - L_{GNK} = L_{G0} - L_{GNK} + 2e_0^0 e_{0,k}^{\hat{0}} U^k + L_{GV} \quad (26)$$

can leads to *the Einstein equations with the character (13) under the condition (3)*, where $L_G$ is given by (4); especially, in $L_{GPK}$, time derivative terms only appear in the term

$$L_{G0} - L_{GNK} = e_{\hat{0}}^{\mu} e_{\hat{0}}^{\nu} \left( \bar{g}^{il} \bar{g}^{jm} - \bar{g}^{ij} \bar{g}^{lm} \right) \Gamma_{\mu ij} \Gamma_{\nu lm} - \frac{2}{3} g^{00} \left[ \frac{1}{2} g^{lm} g_{lm,0} + \frac{1}{2} \frac{g^{0i}}{g^{00}} g^{lm} g_{lm,i} + \left( \frac{g^{0i}}{g^{00}} \right)_{,i} \right]^2 .$$

It is obvious that there is not any negative kinetic energy term in $L_{GPK}$, we therefore can use (25) to quantize gravitation field by various methods of quantization. In this paper, we only discuss the method of canonical quantization.

## 3 The Hamiltonian representation

As a first step of the Hamiltonian representation, we need 3+1 dimensional decomposition of space-time manifold, this can be realized by the well-known ADM decomposition:

$$\mathrm{d}s^2 = -\left(N^2 - h_{ij} N^i N^j\right)(\mathrm{d}x^0)^2 + 2N_i \mathrm{d}x^i \mathrm{d}x^0 + h_{ij} \mathrm{d}x^i \mathrm{d}x^j , \qquad (27)$$

where $N_i = h_{ij} N^j$. For using the form of the foregoing formulas, we still use $g_{ij}$ to denote $h_{ij}$.

Under (3), both (14) and (15) hold in this case, and especially we have

$$e_0^{\hat{0}} = N, \ e_0^{\hat{a}} = e^{\hat{a}i} N_i = e_i^{\hat{a}} N^i, \ e_{\hat{0}}^i = -e_{\hat{0}}^0 N^i, \ h^{ij} = \bar{g}^{ij} = e_{\hat{a}}^i e^{\hat{a}j}. \qquad (28)$$

For the action (25), substituting (21) into (26) and considering (28), we have

$$S = \frac{c^3}{16\pi G} \int \mathrm{d}^4 x \, N h_0 L_{GPK}(h_u; h_{u,\lambda}; N, N_i, h_0; N_{,i}, N_{i,j}, h_{0,i}), \quad u = 1, 2, 3, 4, 5, \qquad (29)$$

defining

$$\pi_u = \frac{\partial (N h_0 L_{GPK})}{\partial h_{u,0}} = \frac{\partial (N h_0 (L_{G0} - L_{GNK}))}{\partial h_{u,0}}, \ u = 1, 2, 3, 4, 5, \qquad (30)$$

and from (30) we obtain $h_{u,0}$ as the functions of $\pi_v$:

$$h_{u,0} = h_{u,0}(\pi_v), \ u, \ v = 1, 2, 3, 4, 5. \qquad (31)$$

Substituting (31) into (29), we have

$$S = \frac{c^3}{16\pi G} \int \mathrm{d}^4 x \, N h_0 L_{GPK}(\pi_u; h_v; h_{v,i}; N, N_i, h_0; N_{,i}, N_{i,j}, h_{0,i}), \quad u, v = 1, 2, 3, 4, 5. \qquad (32)$$

From the above expression we can obtain five constraints:

$$\text{Hamiltonian constraint:} \ \frac{\partial (N h_0 L_{GPK})}{\partial N} - \partial_i \frac{\partial (N h_0 L_{GPK})}{\partial N_{,i}} = 0 , \qquad (33)$$

$$\text{Diffeomorphism constraint:} \ \frac{\partial (N h_0 L_{GPK})}{\partial N_{,i}} - \partial_j \frac{\partial (N h_0 L_{GPK})}{\partial N_{i,j}} = 0 , \qquad (34)$$

$$h_0 \ \text{constraint:} \ \frac{\partial (N h_0 L_{GPK})}{\partial h_0} - \partial_i \frac{\partial (N h_0 L_{GPK})}{\partial h_{0,i}} = 0 . \qquad (35)$$

After realizing quantization, the commutation relations are

$$\begin{array}{l} [h_u(t, \boldsymbol{x}), \pi_v(t, \boldsymbol{x}')] = \mathrm{i}\hbar \delta_{uv} \delta^3(\boldsymbol{x} - \boldsymbol{x}'); [h_u(t, \boldsymbol{x}), h_v(t, \boldsymbol{x}')] = 0; \\ [\pi_u(t, \boldsymbol{x}), \pi_v(t, \boldsymbol{x}')] = 0; \quad u, v = 1, 2, 3, 4, 5. \end{array} \qquad (36)$$

(33) ~ (35) change to five constraint equations for wave function $\Psi[h_u(x)]$, the five equations of motion of operators are

$$\mathrm{i}\hbar \dot{\pi}_u(t, \boldsymbol{x}) = [\pi_u(t, \boldsymbol{x}), H]; \quad u = 1, 2, 3, 4, 5. \qquad (37)$$

In (37),

$$H = \int d^4x \left( \sum_{u=1}^{5} \pi_u h_{u,0}(\pi_v) - Nh_0 L_{\text{GPK}}(\pi_u; h_v; h_{v,i}; N, N_i, h_0; N_{,i}, N_{i,j}, h_{0,i}) \right). \tag{38}$$

All the concrete forms of (29) ~ (38) obtained by computer are complicated.